%% file: main.tex
\documentclass[conf]{new-aiaa}
\usepackage[utf8]{inputenc}

\usepackage{subcaption}
\usepackage{graphicx}
\usepackage{amsmath}
\usepackage{float}
\usepackage[version=4]{mhchem}
\usepackage{siunitx}
\usepackage{longtable,tabularx}
\title{Industrialisation of spectral/hp element method for incompressible, transitional flow around Formula 1 geometries}

\author{Parv Khurana\footnote{CFD Early Stage Researcher, CFD Technology Group, parv.khurana@mclaren.com} and Alexandra Liosi\footnote{CFD Early Stage Researcher, CFD Technology Group, alexandra.liosi@mclaren.com}}
\affil{McLaren Racing Limited, GU21 4YH Woking, UK}
\author{Spencer J. Sherwin\footnote{Head of Department and Professor of Computational Fluid Mechanics, Department of Aeronautics}}
\affil{Imperial College London, SW7 2AZ London, UK}
\author{Julien Hoessler\footnote{Head of CFD Technology, CFD Technology Group},
Adam Swift\footnote{Project Leader, CFD Technology Group}, Athanasios Chatzopoulos\footnote{Project Leader, CFD Technology Group}, Francesco Bottone\footnote{Contractor, CFD Technology Group}}
\affil{McLaren Racing Limited, GU21 4YH Woking, UK}

\begin{document}

\maketitle

\begin{abstract}
This study applies the high-fidelity spectral/hp element method using the open-source Nektar++ framework to simulate the unsteady, transitional flow around complex 3D geometries representative of the Formula 1 industry. This study extends the work on a previously investigated industrial benchmark, the Imperial Front Wing (IFW), derived from the McLaren MP4-17D race car's front wing and endplate design. A combined configuration of the IFW with a wheel in contact with a moving ground in a rolling state is considered, representing the first instance of such a configuration being simulated using higher-order methods. The rolling wheel combined with the IFW (IFW-W) provides the most realistic industrial configuration until now. The spectral/hp element method is applied to this test case to solve the incompressible Navier-Stokes equations, simulating the flow at a Reynolds number of $\mathbf{2.2 \times 10^5}$. Time-averaged results from the unsteady simulation are compared to experimental Particle Image Velocimetry (PIV) data to assess the model's fidelity, offering insights into its reliability for accurately representing key flow characteristics. This research addresses the challenges and requisites associated with achieving diverse levels of flow resolution using the under-resolved DNS/implicit LES approach.
\end{abstract}

\section{Nomenclature}

{\renewcommand\arraystretch{1.0}
\noindent\begin{longtable*}{@{}l @{\quad=\quad} l@{}}
IFW  & Imperial Front Wing \\
IFW-W &    Imperial Front Wing with Wheel \\
CTU  & Convective Time Unit \\
CFL  & Courant–Friedrichs–Lewy \\
$C_p$& Pressure coefficient \\
$C_{p_0}$& Pressure coefficient \\
$C_L$ & Lift coefficient\\
$C_D$ & Drag coefficient \\
$c$   & Chord \\
d$t$ & Time step \\
$Re$ & Reynolds number \\
$h$  & Ride-height \\
$t*$  & Non-dimensional time \\
$L_c$  & Characteristic length \\
$U$  & Far-field velocity
\end{longtable*}}

\input{intro}
\input{case_setup}
\input{results}
\input{conclusion}

% \bibliography{sample}
\bibliography{main}

\end{document}

%% file: intro.tex
\section{Introduction}
\lettrine{D}{ue} to the competitive nature of the motorsport industry and the need to safeguard competitors' intellectual property, there is a scarcity of available geometries and test cases in the public domain, posing challenges for validating numerical methods used in aerodynamic estimations. However, recent efforts have been made to address this challenge with the introduction of the \href{https://data.hpc.imperial.ac.uk/resolve/?doi=6049}{Imperial Front Wing (IFW)}, as an industrial benchmark \citep{Pegrum2006}. IFW derives from the unraced front-wing configuration of the McLaren MP4-17D race car and is publicly available, providing a valuable resource for aerodynamic studies. IFW is a multi-element wing operating in ground effect, generating a complex system of interacting vortices downstream.
 
\begin{figure}[H]
    \centering
    \includegraphics[width=0.8\textwidth]{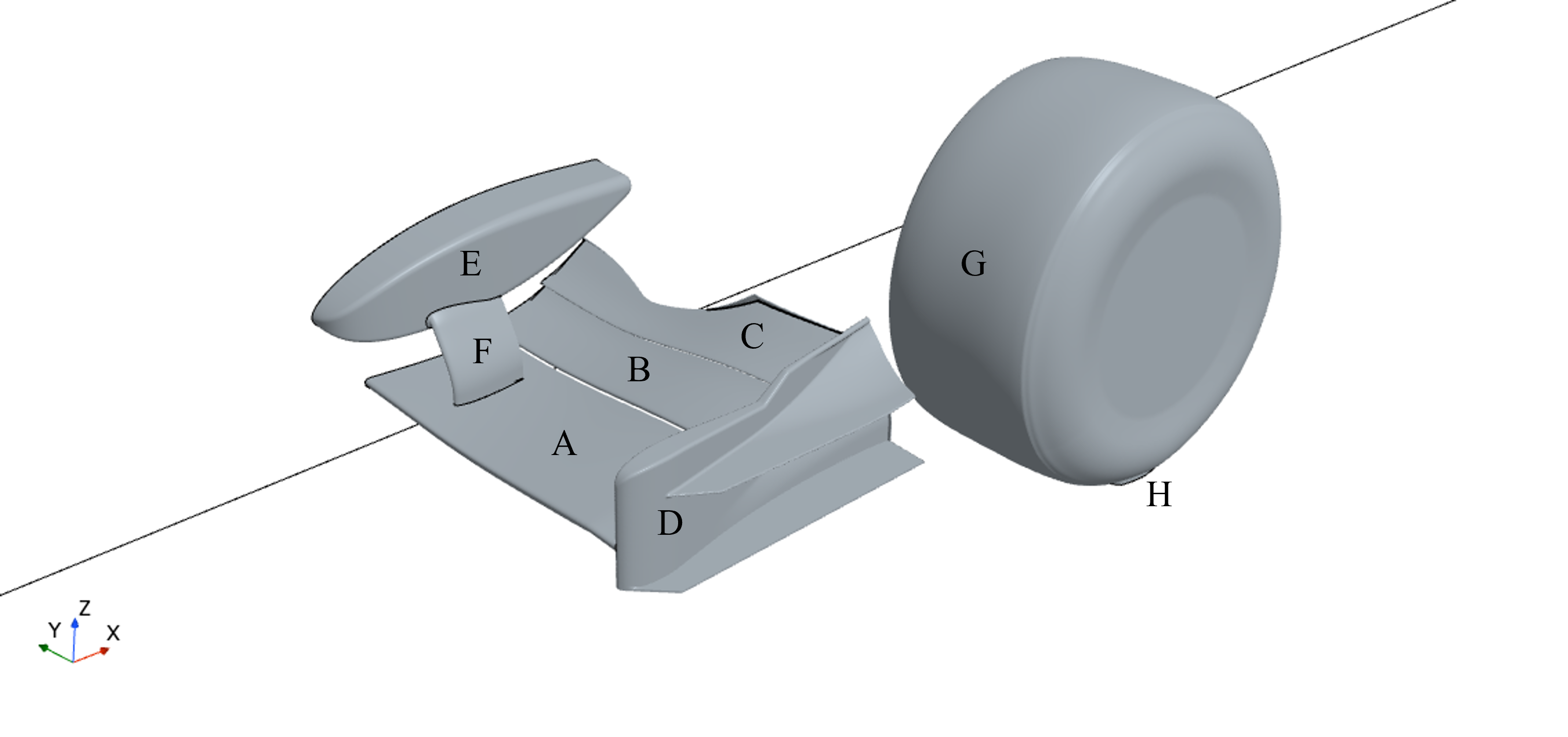} % Replace 'example-image.jpg' with the filename of your picture
    \caption{Primary test case for this study - Imperial Front Wing with a rolling wheel (IFW-W). A: Mainplane, B: First Flap, C: Second Flap,
D: Endplate, E: Nose Cone, F: Hanger, G: Wheel, H: Contact patch}
    \label{fig:IFW_testcases}
\end{figure}

Another industry-relevant test case is an isolated rolling wheel, which in its simplest form is a low aspect ratio bluff body in contact with the ground. This test case is deceptively difficult to understand, with limited data available only for low Reynolds number (Re) flows \citep{fackrell1974aerodynamics, Pirozzoli2012, lombard2017high}. Challenges include modelling the wheel's rotation, moving ground and the contact between the two. Furthermore, it produces a large, unsteady wake and is characterised by separation from the adverse pressure gradient along the circumferences and the rotating edges of the wheel. The combination of the IFW and the rolling wheel (IFW-W), as seen in \autoref{fig:IFW_testcases}, takes us to a more representative model for modern open-wheel motorsports, especially Formula 1 front-wing designs, where the accurate modelling of the vortex systems and their interaction is critical for successful front-wing development. High-fidelity simulations of the IFW and IFW-W present an opportunity to establish a comprehensive database for an industrial benchmark, offering insights that could significantly benefit the automotive/motorsport aerodynamics community.

% \subsection{IFW correlation with PIV data}

Within automotive and motorsport computational workflows, methods like Reynolds-Averaged Navier Stokes (RANS) using Finite Volume (FV) discretisations are widely adopted due to their quick turnaround times and reasonable approximations of critical flow features. However, to understand the intricacies of complex flow dynamics and vortical systems around typical motorsport geometries, a higher-fidelity approach capable of predicting instantaneous flow snapshots is needed. Integrating such a high-fidelity method within an industrial Computational Fluid Dynamics (CFD) pipeline can be a pivotal performance differentiator. This study evaluates the feasibility of incorporating one such high-fidelity method, the spectral/hp element method \citep{karniadakis2005spectral}, for Implicit Large Eddy Simulation (iLES) simulations on complex motorsports geometries. This method has been previously applied to automotive geometries \citep{Hambli2022, Mengaldo2021} and also the IFW \citep{lombard2017high, Buscariolo2022, Slaughter2022}.

The spectral/hp element method provides good numerical convergence properties and offers higher accuracy with fewer degrees of freedom \citep{Moxey2020}  while retaining the geometric flexibility that is salient to traditional linear finite element methods (FEM). Furthermore, the high arithmetic intensity makes them a good candidate for extracting performance from exascale computing systems. Thus, this method is a good candidate for high Reynolds (Re) number flow simulations around large, complex industrial geometries. The open-source software \textit{Nektar++} \citep{Cantwell2015} implements spectral/hp element methods for a variety of flow problems, and it is the chosen software for the study. Typically, the flow around the IFW test case is in a subsonic, incompressible regime; therefore, the focus will be on the incompressible Navier-Stokes equations solver in \textit{Nektar++}. 

This study extends previous work on incompressible flow simulations for the Imperial Front Wing (IFW) using Nektar++ for a more complex open-wheel Formula 1 geometry. The combined configuration of the IFW and the rolling wheel (IFW-W) is investigated. It is the first application of a higher-order method to this geometry. Detailed numerical setups for these cases are outlined, offering recommendations on meshing strategies, solver setup, and solution restart methodologies, focusing on streamlining these processes for industrial applicability. Computational performance is assessed, elucidating resource requirements for varying levels of fidelity. Validation of the methodology is conducted by comparison of time-averaged simulation results with experimental data for the IFW from \citet{Buscariolo2022}, ensuring the reliability of the computational models. Following validation, an analysis is undertaken to determine the acceptable level of flow resolution that maintains solution accuracy while expediting high-order simulations. Based on this evaluation, a set of guidelines is proposed, marking a step towards industrializing the spectral/hp element approach for motorsport CFD workflows.

%% file: case_setup.tex
\section{Case Description}
This section describes the steps and tools to run spectral/hp element method simulation on the selected test cases. The simulations are run for a moderate Reynolds number $Re = 2.2 \times 10^5$, based on previous studies \citep{Pegrum2006, Buscariolo2022, lombard2017high, Slaughter2022}. 

\subsection{Incompressible Navier-Stokes solver using spectral/hp element method}
The Incompressible Navier-Stokes (IncNS) solver for viscous Newtonian fluids is implemented in \textit{Nektar++}. The IncNS equations can be written as follows: 
\begin{subequations}
\begin{equation}
   \frac{\partial \mathbf{u}}{\partial t} + \mathbf{u} \cdot \nabla \mathbf{u} = -\nabla p + \nu \nabla^2 \mathbf{u} +  \mathbf{f} \label{eqn.NSmom}
 \end{equation}
 \begin{equation}
    \nabla \cdot \mathbf{u} = 0
    \end{equation}
    \label{eqn:IncNS}
\end{subequations}
where $\mathbf{u}$ is the velocity, $p$ is the specific pressure (including density), $f$ is the forcing term and $\nu$ the kinematic viscosity. The IncNS solver in \textit{Nektar++} uses a Velocity Correction Scheme (VCS) or a high-order splitting scheme \citep{KARNIADAKIS1991414}, which decouples the velocity and the pressure system.

Using the spectral/$hp$ element method to discretise the governing equations \autoref{eqn:IncNS}, the dependent variable is represented using the following expansion in terms of the elemental modes:
\begin{equation}
  u^{\delta}(\mathbf{x})=\sum_{e=1}^{{N_{\mathrm{el}}}}
  \sum_{n=0}^{N^{e}_m-1}\hat{u}_n^e\phi_n^e(\mathbf{x})
  \label{eqn:precon:disc}
\end{equation}
where the domain is decomposed into $N_{\mathrm{el}}$ total elements, where $\Omega^e$ signifies an element. $N^{e}_m$ denotes the local polynomial expansion modes within the element $\Omega^e$, $\phi_n^e(\mathbf{x})$ signifies the $n^{\mathrm{th}}$ local expansion mode within the element $\Omega^e$, and $\hat{u}_n^e$ represents the $n^{\mathrm{th}}$ local expansion coefficient within the element $\Omega^e$. The approximation in \autoref{eqn:precon:disc} is utilized to adopt classical continuous Galerkin discretisation of the equation systems arising from splitting \autoref{eqn:IncNS}.

\subsection{Mesh generation}
The objective of generating industrial meshes for spectral/hp element methods is to construct coarser meshes than their finite volume counterparts, leveraging higher-order polynomials to achieve accuracy with fewer degrees of freedom. Creating higher-order meshes of complex geometries in the current framework involves two steps: (a) generating a linear mesh and (b) converting it to a high-order curvilinear mesh. The resulting mesh is a conformal, unstructured, mixed-element mesh, featuring prismatic elements in the near-surface boundary layer and tetrahedral elements in the far-field domain.

\begin{figure}[H]
    \centering
    \includegraphics[width=0.8\textwidth]{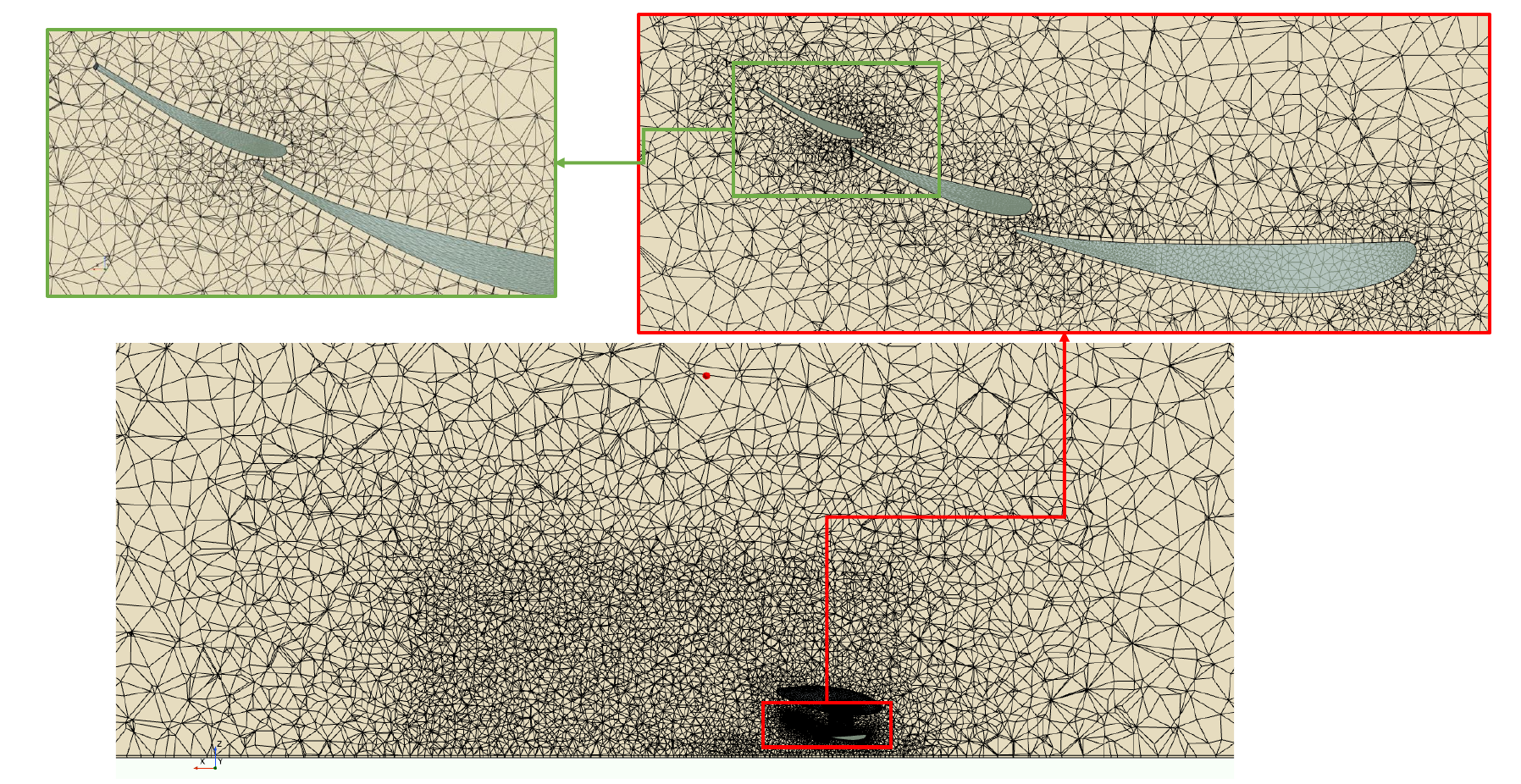}
    \caption{Slice at Y = -0.250 m for the linear mesh generated for the IFW-W case. The refinement boxes can be identified around the geometry. The macro prism layer is visible around the airfoil elements and on the floor.}
    \label{fig:IFWW_meshy250}
\end{figure}

\begin{figure}[H]
    \centering
    \includegraphics[width=0.9\textwidth]{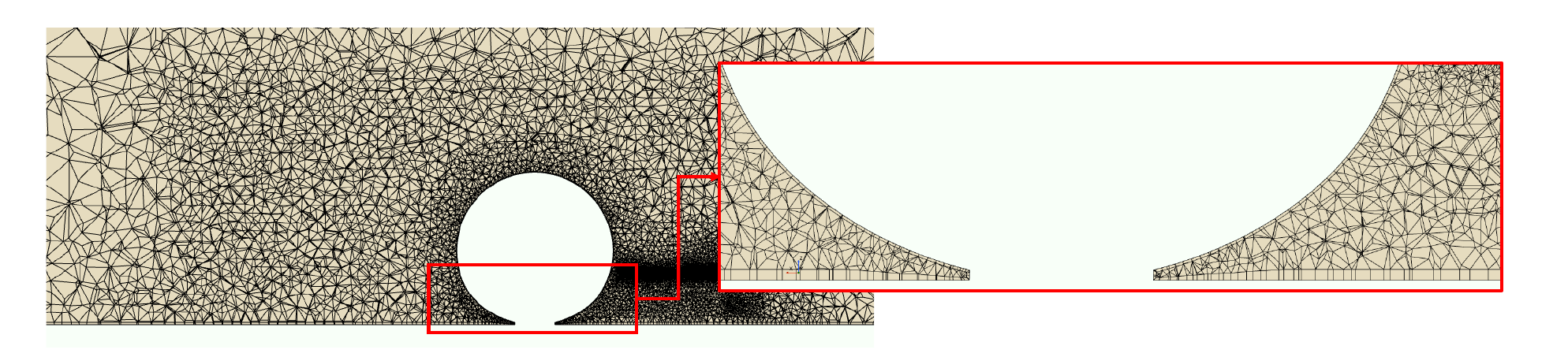} 
    \caption{Slice at Y = -0.7118 mm for the linear mesh generated for the IFW-W case. The wheel is surrounded by a macro prism layer which shrinks near the contact patch, and then matches the thickness of the floor.}
    \label{fig:IFWW_meshy7118}
\end{figure}

The first step, or the linear mesh, is generated using commercial Finite Volume meshing software and has tetrahedron and prismatic elements and a single "macro" prism layer. The single "macro" prism layer is a requirement for the linear mesh, with the first cell height being the total desired thickness of the boundary layer mesh. The surface mesh is first generated based on specified user settings. The volume mesh is then propagated from the surface with a Prism layer Mesher, which generates the "macro" prism layer. The Tetrahedral layer Mesher then generates tetrahedrons in the remaining domain. Cross-sections of the linear mesh in \autoref{fig:IFWW_meshy250} and \autoref{fig:IFWW_meshy7118} show these features in the resulting linear mesh. The first cell height is 2.5 mm for the IFW-W geometry and 10 mm for the floor to capture the near-wall physics. Appropriate volume refinement is applied to capture the flow in the wake of the geometry. A gradual transition from one refinement box to another is advised as it aids the stability of the solution.
\vspace{-5mm}

% \subsection{IFW correlation with PIV data}
\begin{figure}[H]
    \centering
    \includegraphics[width=\textwidth]{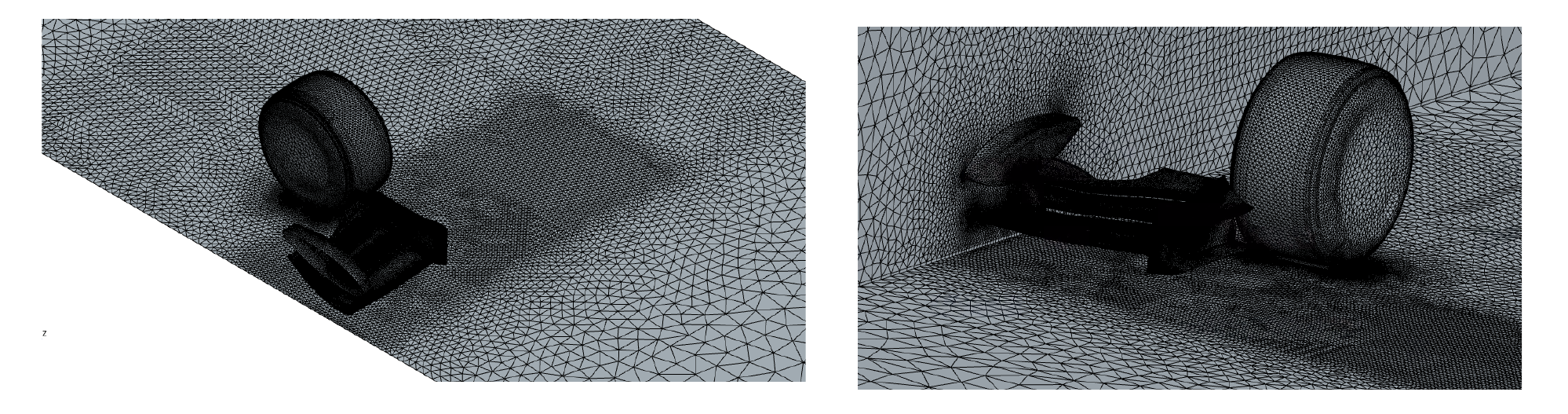} % Replace 'example-image.jpg' with the filename of your picture
    \caption{Linear mesh generated for the IFW-W case. The surface mesh is visible over the wing elements and the wheel.}
    \label{fig:IFW-Wmesh}
\end{figure}
For the second step, the linear mesh is processed by the higher-order mesh generation utility of Nektar++ - \textit{NekMesh} \citep{green4632754nekmesh}. This utility does the mesh projection step on the CAD boundary and adds higher-order nodes on edges, faces, and elemental interiors. The CAD projection order is chosen as 4 to reconstruct the CAD information for this complex geometry accurately. This restricts the polynomial order of the simulation as it has to be greater than or equal to CAD projection order to avoid geometrical aliasing \cite{mengaldo2015dealiasing}.    

The macro prism layer is then split using an iso-parametric approach to refine the boundary layer to the desired first cell height and distribution. After the splitting, the final number of prism layers is set as 8 for the IFW-W geometry and 5 for the floor, with a growth rate of 1.2 to capture the near-wall physics. More details on the higher-order mesh-generation process from third-party linear meshes for industrial cases in Nektar++ can be found in \citet{green4632754nekmesh}.

\begin{figure}[H]
\centering
\includegraphics[width=0.5\linewidth]{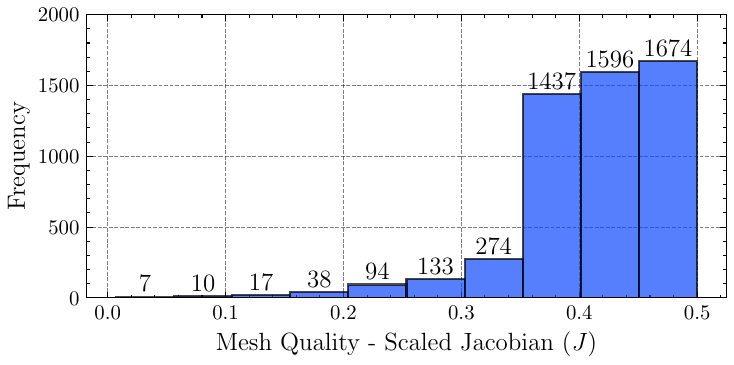}
\caption{Histogram of the scaled Jacobian < 0.5 of all the mesh elements. Total mesh elements: $\mathbf{2.87 \times 10^6}$}
\label{fig:IFWW_mesh_histo}
\end{figure}

To evaluate the mesh quality of the higher order mesh, \textit{NekMesh} uses an element-wise quality measure chosen: the scaled Jacobian of the mapping between the ideal and curved element, where a value of 1 denotes an ideal element with no curve, and less than zero indicates invalid elements. \autoref{fig:IFWW_mesh_histo} shows the distribution of scaled Jacobian of all the mesh elements until < 0.5, focusing on the bad-quality elements.

\begin{figure}[H]
    \centering
    \includegraphics[width=\textwidth]{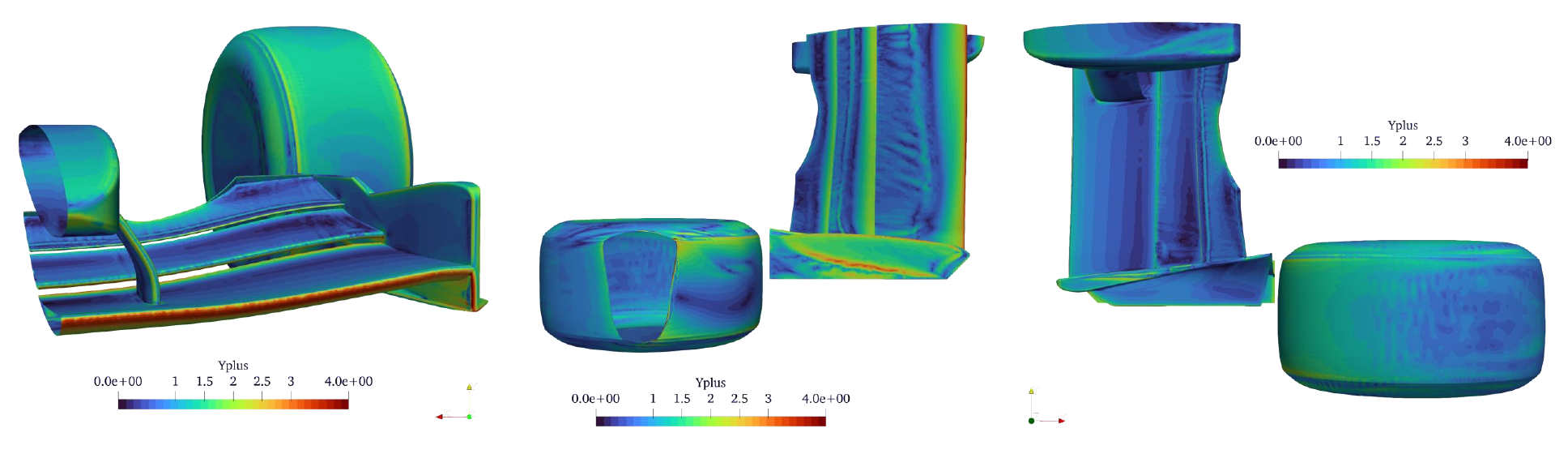} % Replace 'example-image.jpg' with the filename of your picture
    \caption{Mesh resolution in wall units $y+$ for the averaged solution for the last 4 CTUs of the IFW-W simulation. }
    \label{fig:IFWW_yplus}
\end{figure}

The above metric indicates mesh quality, but it is challenging to know a-priori whether a good mesh will resolve sufficient flow physics. Near-wall physics can be verified using mesh resolution in terms of wall units, i.e. $y^+ = \frac{u_{\tau} * y}{\nu}$, where $u_{\tau}$ is the wall shear stress magnitude, and $\nu$ is the kinematic viscosity. For traditional lower-order discretisation, the $y$ is the first cell height. However, for higher-order meshes considered here, $y$ is the position of the first quadrature point within the element. For ease of calculation, the quadrature points are assumed to be spaced equidistant; thus, the resulting $y^+$ overestimates the actual value. \autoref{fig:IFWW_yplus} gives the $y^+$ values over the surface of the geometry for an averaged solution of the IFW-W simulation. The limits of resolution for wall-resolved LES are presented by \citet{georgiadis2010large}.

\subsection{Case and Solver setup}
The chosen $Re = 2.2 \times 10^5$ is based on the characteristic length chosen as the main plane chord of the IFW, i.e. $L_c= 0.25m$. A non-dimensional time $t^* = t U/ L_c$ is based on the far-field velocity $U$ and characteristic length $L_c$, and one unit of this non-dimensional time is referred to as Convective Time Unit (CTU) in this study. The default chosen ride height ($h$), which is the distance between the ground and the lowermost point of the car, is $0.36 h/L_c$ for this case, where the height is measured from the footplate of the full geometry. The dimensions of the computational domain can be seen in \autoref{fig:DomainBCs}, designed to emulate a wind tunnel with a rolling road.

The polynomial order range for this study varies between 3 and 5, following previous industrial applications of Nektar++ \citep{Hambli2022, Buscariolo2022, Slaughter2022}. The choice of polynomial order is made using a Taylor–Hood approximation to satisfy the inf-sup condition \cite{bathe2001inf}. This means that the polynomial order of the expansion for pressure will be one less than that of the velocity. Thus, $P=4,3$ denotes the polynomial order of 4 for the velocity field and 3 for the pressure field. Furthermore, more quadrature points are used for each expansion to ensure exact integration and avoid polynomial aliasing \cite{mengaldo2015dealiasing}. Simulation parameters are normalized in Nektar++. Therefore, far-field velocity is $U = 1 m/s$, and the viscosity is scaled according to the $Re$.
% -Add SVV stability
% -Dealiasing
% -Static condensation
% -time-stepping order
% -CG solver
% -Preconditioning
% -Solver tolerances

The IncNS solver framework in Nektar++ uses the VCS, a semi-implicit scheme in which the diffusion terms are treated implicitly, and the advection terms are handled explicitly. The chosen time step (d$t$) to run the simulation is bounded by the simulation's stability, governed by the Courant–Friedrichs–Lewy (CFL) criterion imposed by the explicit treatment of advection terms. Time integration is performed using a second-order accurate implicit-explicit (IMEX) scheme, ensuring a balance between accuracy and computational efficiency. Stabilization is achieved using the Spectral Vanishing Viscosity DG Kernel \citep{rosh06}. The discretized system of equations is solved iteratively using the Conjugate Gradient (CG) algorithm, with static condensation applied to the system matrix \citep{karniadakis2005spectral} to reduce the number of algebraic degrees of freedom. Algebraic convergence for each equation is assessed using an absolute tolerance criterion, requiring the algebraic error for every flow variable to be below $10^{-4}$. The velocity system is preconditioned using the LowEnergyBlock approach, based on the methodology by \citet{SherwinCasarin2001}. The pressure system is preconditioned using the standard Diagonal preconditioner. However, more sophisticated approaches based on lower-order discretisations are being explored \cite{khurana2023comparison, khurana2024LOR}.

%     \item \textbf{LowEnergyBlock}: This approach is based on the methodology by \citet{SherwinCasarin2001}. A new expansion basis is generated from the original by choosing a suitable transformation matrix. Then, a Schur complement matrix of the matrix system with the new basis is generated, which is then preconditioned using the block preconditioner. The transformation matrix aims to make the original matrix system amenable to block preconditioning. 

\subsection{Boundary conditions}
\begin{figure}[H]
    \centering
    \includegraphics[width=0.9\textwidth]{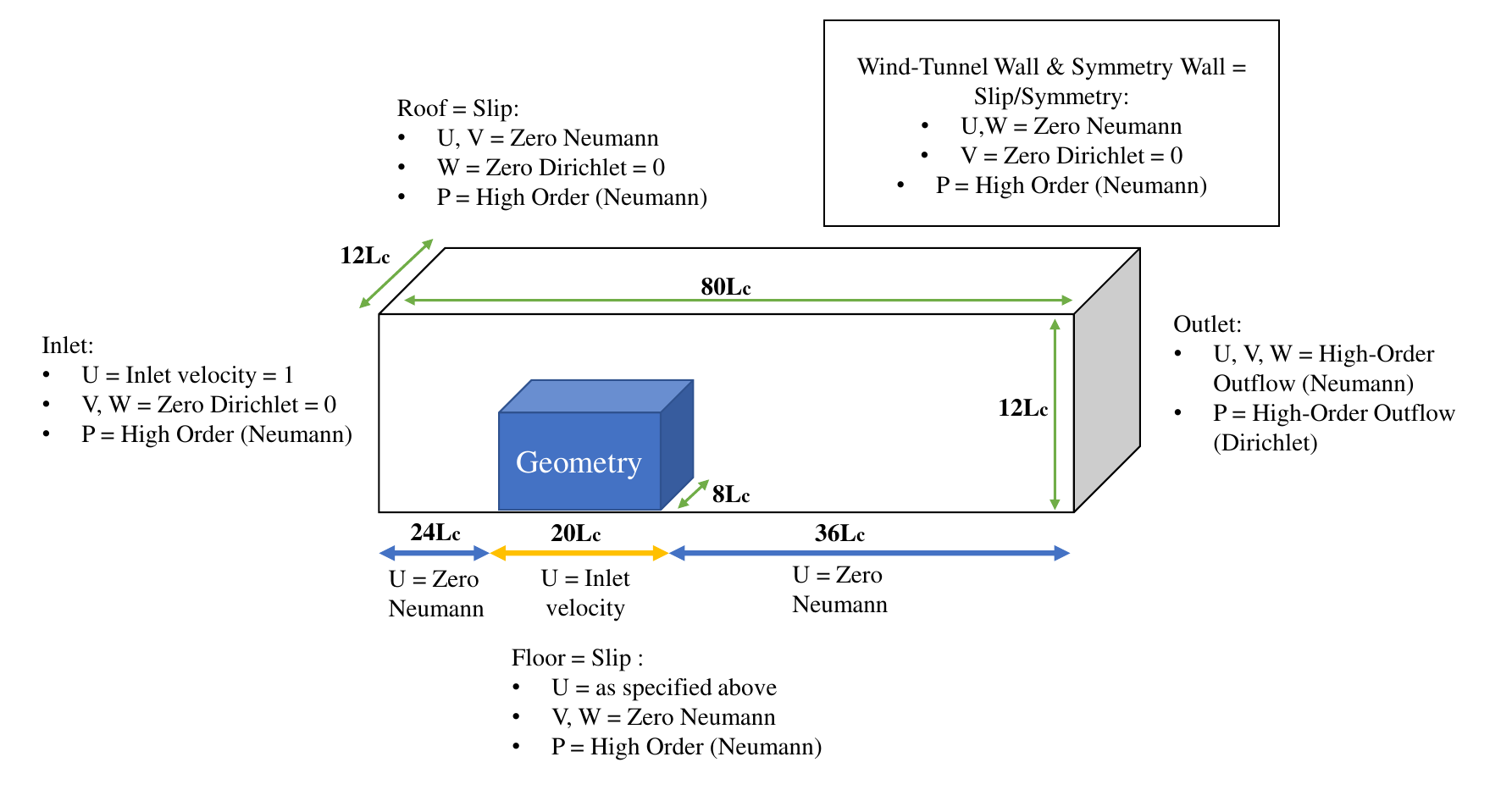}
    \caption{Schematic for the boundary conditions applied on the computational domain used for this study. The domain dimensions are not to scale. This setup is used for the IFW and IFW-W simulations}
    \label{fig:DomainBCs}
\end{figure}

A uniform inflow is imposed at the inlet in a Dirichlet manner. The floor consists of three components: a moving wall with a velocity like the inflow to replicate the rolling road in a wind tunnel. This moving wall is surrounded by two patches of stationary ground, with the slip condition imposed for the velocity. The slip condition is also applied to the ceiling of the domain. The side walls have the symmetry condition imposed, allowing zero cross-flow. For pressure, a high-order Neumann boundary condition  \citep{KARNIADAKIS1991414, GuermondandShen2003} is applied to all the computational domain walls apart from the outlet. For the outlet, a high-order Dirichlet convective boundary condition is applied \citep{DoKa14}. The boundary conditions for the computational domain can be seen in \autoref{fig:DomainBCs}. 
The boundary conditions for IFW-W geometry are no-slip, i.e. the pressure is zero Neumann, and all velocity components are Dirichlet. The wing has a zero Dirichlet specified on all surfaces. For the wheel, the velocity components are imposed in a Dirichlet manner by calculating the rotational velocity at every point:
\begin{equation}
\vec{\boldsymbol{v}} = \vec{\boldsymbol{\omega}} \times \vec{\boldsymbol{r}}, \quad \text{where} \quad \vec{\boldsymbol{r}} = \begin{bmatrix} x - x_c \\ y - y_c \\ z - z_c \end{bmatrix}, \quad \vec{\boldsymbol{\omega}} = \begin{bmatrix} \omega_x \\ \omega_y \\ \omega_z \end{bmatrix},
\end{equation}
$(x_c, y_c, z_c)$ is the centre of rotation of the wheel, and $(\omega_x, \omega_y,\omega_z)$ are the rotational vector components of the wheel. The wheel's contact patch can be modelled in two ways: by imposing a Dirichlet boundary condition for the rotational velocity or by specifying the velocity components to match the rolling road. This study adopts the latter approach, providing more physically accurate pressure distributions around the contact patch. However, a more detailed analysis of the contact patch physics falls outside the scope of the current work.

\subsection{Running the simulations}
% -Restart from RANS
% -Raise the polynomial order ASAP, then push timestep until the stability limit
% -Start with more strict tolerance, then relax to 1e-4 for pressure and 1e-2 for velocity
The simulation is restarted from an initial solution obtained via a RANS model run on the same geometry, with the velocity field taken from the solution and pressure set to zero. Restarting from uniform inflow conditions is considered infeasible due to several factors: timestep restrictions at the simulation start, the time required to advance the solution to the geometry, and the overly "clean" nature of the uniform conditions, which hinder the development of turbulent flow. Alternative restart strategies, such as initialising from a turbulent inflow generator or employing a more unsteady solution like a DES simulation, have not been explored in this study.

\begin{figure}[H]
\centering
\includegraphics[width=0.8\linewidth]{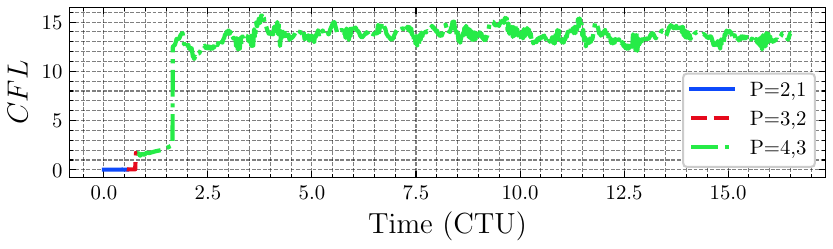}
\caption{Evolution of max CFL value at every solver time step. The variation of polynomial order is indicated with different colours. The timestep was increased to the maximum stable value of d$t$ = $5 \times 10^{-6}$ sec at $t^* \approx 1.6$.}
\label{fig:IFWW_CFL}
\end{figure}

The current IFW-W simulation starts with $P=2,1$ and a d$t = 1 \times 10^{-6}$ sec. The CFL criterion imposes an upper bound on the time step. Once the simulation runs stably, the polynomial order is increased in increments of one. The idea is to achieve the best spatial discretisation as early as possible in the simulation to avoid spatial errors propagating in the solution. Once the desired polynomial order is achieved, the time step is progressively increased until the stability limit of the simulation is reached. The simulation is then continued to run at that final configuration. The process can be seen in \autoref{fig:IFWW_CFL}, where the maximum CFL in the domain is plotted with a progressing solution in time. The IFW-W simulation finally runs at $P=4,3$ and a d$t = 5 \times 10^{-6}$ sec.  

Complex geometries, such as the IFW-W, exhibit a wide range of spatial scales, from small corners ($\approx 10^{-3}$ m) to large wings ($\approx 1$ m). The flow velocity and the smallest geometry dimensions primarily govern the timestep restrictions. However, the small mesh elements with bad quality are typically localized in regions that either have minimal impact on the significant flow structures or are not critical to the overall stability of the simulation. 

% Top standalone figure
\vspace{-4mm}
\begin{table}[H]
\centering
\begin{tabular}{lcccl}
\hline
\textbf{Element ID} &
  \multicolumn{1}{l}{\textbf{Max CFL Count}} &
  \multicolumn{1}{l}{\textbf{\% of total steps}} &
  \multicolumn{1}{l}{\textbf{Scaled Jacobian}} &
  \textbf{Element type} \\ \hline
\textbf{448143}  & 108943 & 85.22\% & 0.0179413        & Prism \\
\textbf{537175}  & 15247  & 11.93\% & 0.0114746         & Prism \\
\textbf{1377160} & 1433   & 1.12\% & 0.0459862          & Prism \\
\textbf{613176}  & 638    & 0.50\% & 0.472105           & Prism \\
\textbf{1048252} & 512    & 0.40\% & > 0.5  & Prism \\ \hline
\end{tabular}
\caption{Mesh element IDs with the frequency of occurrences as the maximum CFL at each solver step, along with their quality metrics.}
\label{tab:IFWW_mesh_CFL}
\end{table}

As shown in \autoref{fig:IFWW_CFL}, the simulation runs stably with a maximum instantaneous CFL value of $\approx 15$ in the domain. The element with the maximum occurrences of the highest CFL (ID = 448143, as detailed in \autoref{tab:IFWW_mesh_CFL}) is located at the junction between the endplate and the first flap. This element also exhibits a poorly scaled Jacobian value close to zero. The ability of the simulation to run stably at CFL >> 1 is because the critical regions of the domain remain governed by the CFL stability criterion, with the outlier element(s) skewing the maximum CFL indicator. Within the current framework, it is recommended to push the timestep as much as possible for industrial simulations, testing the CFL limits beyond unity to determine the actual stability threshold. 

The iterative solver is initially run with stricter tolerances of $10^{-4}$ for both pressure and velocity systems to ensure stability and accuracy during the early stages. As the simulation progresses, these tolerances are relaxed to $10^{-2}$ for the velocity system to accelerate the solution time of the simulation. If the pressure system is solved with a tolerance of $10^{-4}$ and lower, the velocity system can be solved with a loose tolerance without introducing significant algebraic error in the solution.

% % Second row: figure and table side by side
% \begin{figure}[H]
% \centering
% \begin{minipage}[t]{0.4\linewidth}
%     \centering
%     \raisebox{-1.7cm}{
%     \includegraphics[width=\linewidth]{Pictures/IFWW_mesh_histo.pdf}
%     }
%     \caption{Mesh element histogram with max CFL values.}
%     \label{fig:IFWW_mesh_histo}
% \end{minipage}%
% \hspace{0.5cm} % Horizontal space between figure and table
% \begin{minipage}[t]{0.55\linewidth}
%     \centering
%     \resizebox{\linewidth}{!}{%
%     \begin{tabular}{lcccl}
%     \hline
%     \textbf{Element ID} & \textbf{Count} & \textbf{Scaled Jacobian} & \textbf{\% Occurrence} & \textbf{Type} \\ 
%     \hline
%     448143  & 108943 & 0.0179413        & 85.22\% & Prism \\
%     537175  & 15247  & 0.0114746        & 11.93\% & Prism \\
%     1377160 & 1433   & 0.0459862        & 1.12\%  & Prism \\
%     613176  & 638    & 0.472105         & 0.50\%  & Prism \\
%     1048252 & 512    & > 0.5 & 0.40\%  & Prism \\ 
%     \hline
%     \end{tabular}%
%     }
%     \captionof{table}{Mesh element IDs with max CFL and their quality metrics.}
%     \label{tab:IFWW_mesh_CFL}
% \end{minipage}
% \end{figure}

%% file: results.tex
\section{Results}
This section presents the simulation results obtained using the industrial setup detailed in the previous section. First, the current method is validated by reproducing the results for IFW by \citet{Buscariolo2022}. Subsequently, the results for the IFW-W are presented. The analysis begins by assessing the convergence of the unsteady simulations by interpreting various metrics and their applicability in an industrial environment. The discussion focuses on identifying practical considerations for optimizing simulation performance while maintaining the fidelity required for industrial applications. The simulations are run on computational nodes equipped with dual AMD EPYC 7742 processors, each with 64 cores operating at 2.25 GHz, fully utilized. Each node is configured with 256 GB of memory, and the nodes communicate via the HPE Cray Slingshot interconnect, providing a bi-directional bandwidth of 2 × 100 Gbps per node.

\subsection{Validation with previous studies}
The IFW geometry is simulated for a polynomial order of $P=4,3$. The simulation is time-averaged based on two criteria: a) the moving standard deviation
of the force traces of the geometry to be below 1\% and b) the temporal derivative of the sampled window (1 CTU) for the force traces be lower than $10^{-6}$. The comparison of the force coefficients in \autoref{tab:IFWresults} already shows a good agreement between the current results and \citet{Buscariolo2022}.
\begin{figure}[H]
    \centering
    \includegraphics[width=0.85\textwidth]{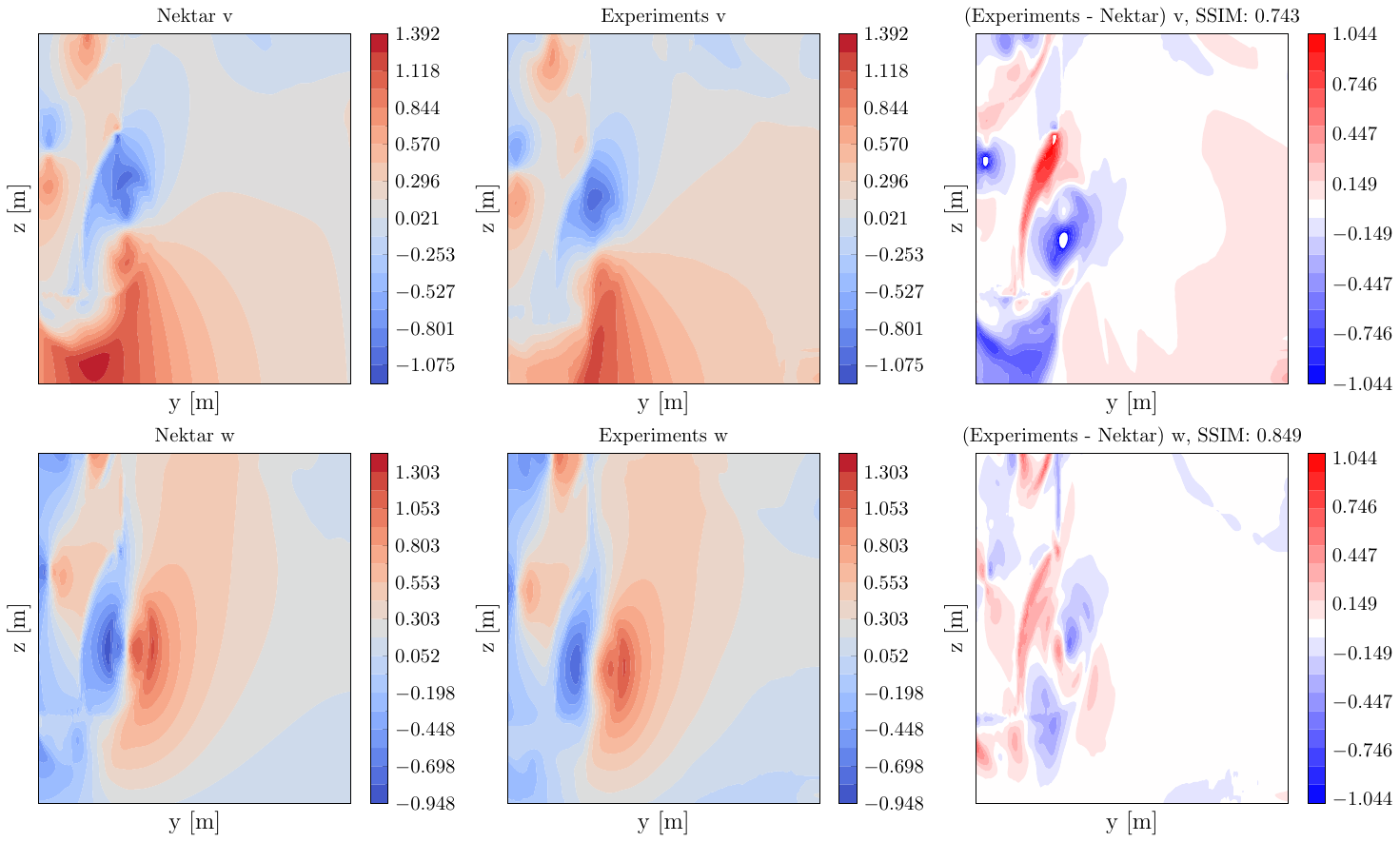} % Replace 'example-image.jpg' with the filename of your picture
    \caption{Comparison of the averaged velocity components between the Nektar++ results (left) and the experimental PIV data (right) for averaged data at the location of PIV plane 1 from \citet{Buscariolo2022}. The SSIM is the structural similarity index \cite{Wang2004} used to quantify the correlation between the results.}
    \label{fig:PIVplanes}
\end{figure}
To validate the development of flow features for the IFW, the current time-averaged IFW results are compared with the available experimental PIV plane data from \citet{Buscariolo2022}. The comparison of the spanwise and normal velocity components for PIV plane 1 is seen in \autoref{fig:PIVplanes}, where the main vortical structures for the current study qualitatively agree with the experiments. For a more detailed discussion of the validation, the reader is referred to the work by \citet{liosi2024DLES}.
\begin{table}[H]
\caption{Comparison of the time-averaged force coefficients with \citet{Buscariolo2022}}
\label{tab:IFWresults}
 \centering
 \begin{tabular}{ccccccc}
 \hline
Scheme & timestep, sec & $C_L$ & $C_D$ & $C_L$ Difference, $\%$ & $C_D$ Difference, $\%$\\\hline
Reference &  & -5.540 & 0.575 & N/A & N/A \\
Current results & $5 \times 10^{-6}$ sec &  $-5.676 \pm 0.001$  & $0.593 \pm 0.002$ & 2.24 & 2.87 \\
 \hline
\end{tabular}
\end{table}
\subsection{IFW-W simulations}
The IFW-W simulations are run for a polynomial order of $P=4,3$. The computational performance statistics for the final 1.5 Convective Time Units (CTUs) are summarized in \autoref{tab:archer2_stats}. The simulations is run on 60 computational nodes, with an average runtime of 43.2 hours per CTU, as detailed in \autoref{subtab:simulation_stats}. For a statistically converged solution, it is generally recommended to average at least five CTUs of unsteady simulation output to capture the flow phenomena adequately. Under the current configuration, approximately nine days of computation are required to collect this information. The pressure system takes approximately 27\% of the execution time, compared to 60\% for all velocity components combined. The pressure system also shows a high average iteration count with a large standard deviation. 
\begin{table}[H]
    \caption{Simulation statistics for an IFW-W run at $P = 4,3$ and $dt = 5 \times 10^{-6}$ sec}
    \begin{subtable}{.4\linewidth}
      \centering
        \caption{Timings, Computation resources}
        \label{subtab:simulation_stats}
        \begin{tabular}{ll}
            \hline
            \textbf{Property}               & \textbf{Value} \\ \hline
            \textbf{MPI ranks} & 60 * 128  = 7680                \\
            \textbf{Linear mesh elements}   & $2.87 \times 10^6$ \\
            \textbf{Total time steps ran}   & 75000          \\
            \textbf{Avg Time per iteration} & 3.11 sec       \\
            \textbf{Time to CTU}            & 43.2 hrs       \\ \hline
        \end{tabular}
    \end{subtable}%
    \begin{subtable}{.6\linewidth}
      \centering
        \caption{Iterative solver statistics over 75000 steps}
        \label{subtab:presure_velocity_stats}
        \begin{tabular}{lll}
            \hline & \textbf{Velocity \textit{P}=4} & \textbf{Pressure \textit{P}=3}   \\
            \hline
            \textbf{Degrees of Freedom} & $76.1 \times 10^6$ & $32.2 \times 10^6$ \\
            \textbf{Preconditioner} & LowEnergyBlock         & Diagonal               \\
            \textbf{Solver tolerance}                 & $10^{-2}$ & $10^{-4}$ \\
            \textbf{Mean CG iterations}           & $70 \pm 0.5$                     & $322 \pm 461$                    \\
            \textbf{\% of execute time}     &  60.35                 &  27.10                  \\ \hline
        \end{tabular}
    \end{subtable} 
    \label{tab:archer2_stats}
\end{table}

\begin{figure}[H]
\centering
\begin{subfigure}{0.5\textwidth}
  \centering
  \includegraphics[width=\linewidth]{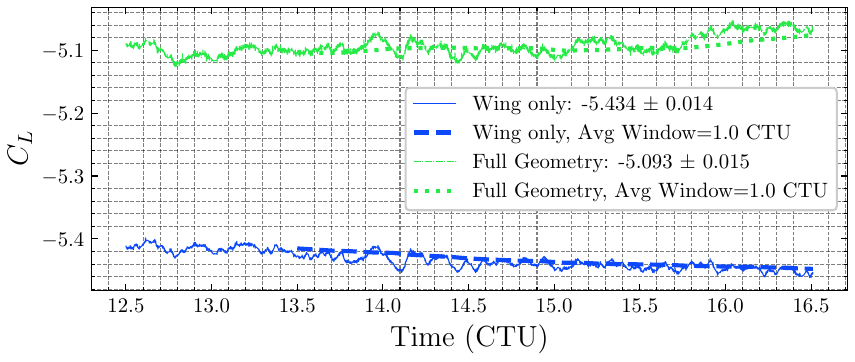}
  \caption{Lift Coefficient $C_L$}
  \label{subfig:IFWW_CL}
\end{subfigure}%
\begin{subfigure}{0.5\textwidth}
  \centering
  \includegraphics[width=\linewidth]{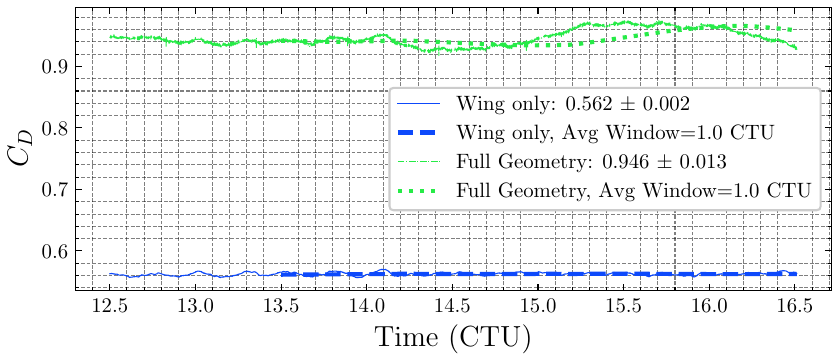}
  \caption{Drag Coefficient $C_D$}
  \label{subfig:IFWW_CD}
\end{subfigure}
\caption{Evolution of the force coefficients for Wing only (A+B+C+D from \autoref{fig:IFW_testcases}) and the Full Geometry for the last 4 CTUs of the simulation}
\label{fig:IFWW_force_coefficients}
\end{figure}

\begin{table}[H]
\centering
\caption{Averaged force coefficients and max temporal derivative of the force signals over a sample window of 1 CTU for the IFW-W simulations.}
\label{tab:forces_IFWW}
\resizebox{\textwidth}{!}{%
\begin{tabular}{llcccc}
\hline
 &
  \textbf{\autoref{fig:IFW_testcases} parts} &
  \multicolumn{1}{l}{\textbf{$C_L$}} &
  \multicolumn{1}{l}{\textbf{Max derivative $C_L$}} &
  \multicolumn{1}{l}{\textbf{$C_D$}} &
  \multicolumn{1}{l}{\textbf{Max derivative $C_D$}} \\ \hline
\textbf{Full Geometry}         & A+B+C+D+E+F+G+H & $-5.093 \pm 0.015$ & $1.21 \times 10^{-5}$ & $0.946 \pm 0.013$ & $9.97 \times 10^{-6}$ \\
\textbf{Wing only}             & A+B+C+D         & $-5.434 \pm 0.014$ & $8.22 \times 10^{-6}$ & $0.562 \pm 0.002$ & $2.35 \times 10^{-6}$ \\
\textbf{Wheel}                 & G+H             & $0.310 \pm 0.024$  & $1.07 \times 10^{-5}$ & $0.382 \pm 0.013$ & $1.04 \times 10^{-5}$ \\
\textbf{Wing +Nosebox +Hangar} & A+B+C+D+E+F     & $-5.403 \pm 0.014$ & $8.40 \times 10^{-6}$ & $0.564 \pm 0.002$ & $2.43 \times 10^{-6}$\\ \hline
\end{tabular}%
}
\end{table}
The IFW-W solution has been averaged for the last 4 CTUs from $t^* = 12.5$ to $t^* = 16.5$ for further analysis. \autoref{tab:forces_IFWW} presents the averaged forces over the last 4 CTUs and the maximum value of the moving temporal derivative calculated using a sample window of 1 CTU. The corresponding \autoref{fig:IFWW_force_coefficients} illustrates the force evolution during this period. Among the components, the wing accounts for the largest contribution to the lift, while both the wing and the wheel significantly influence the drag. The wing is near convergence according to the criteria used for IFW in the previous section. In contrast, the wheel exhibits a higher standard deviation in this window, indicating greater variability in its force contributions. Notably, the nosebox and hanger do not contribute to the lift.

\begin{figure}[H]
\centering
\begin{subfigure}{0.45\textwidth}
  \centering
  \includegraphics[width=\linewidth]{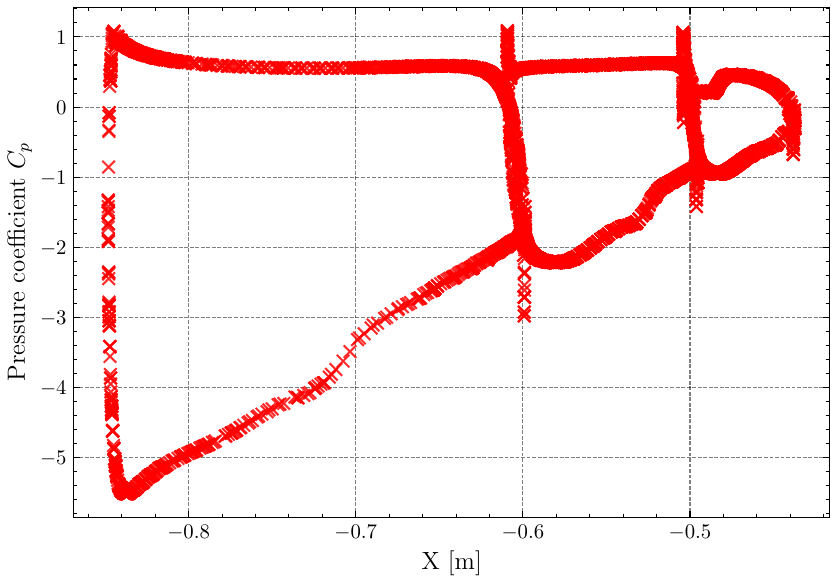}
  \caption{Pressure Coefficient ($C_p$)}
  \label{subfig:IFWW_Cp_y250}
\end{subfigure}%
\hspace{0.5cm}
\begin{subfigure}{0.45\textwidth}
  \centering
  \includegraphics[width=\linewidth]{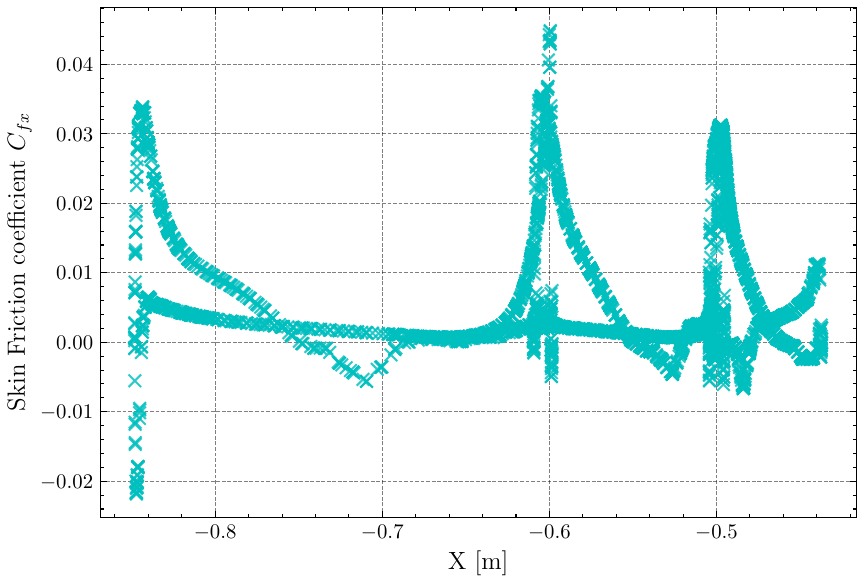}
  \caption{Skin Friction Coefficient ($C_{f_x}$)}
  \label{subfig:IFWW_Cfx_y250}
\end{subfigure}
\caption{Averaged coefficients at a y = -0.250 m slice for the IFW-W simulations}
\label{fig:IFWW_y250}
\end{figure}

The pressure ($C_p$) and the skin friction ($C_{f_x}$) coefficients are evaluated for the time-averaged flow field and plotted on a cross-section at $ y = -0.25$ m in \autoref{fig:IFWW_y250}. Transition regions are identified on the main plane and the first flap at approximately $x = -0.75$ m and $x = -0.54$ m, respectively, as shown in \autoref{subfig:IFWW_Cp_y250}. From the 
$C_{f_x}$ distribution in \autoref{subfig:IFWW_Cfx_y250}, it is observed that flow separation occurs at these locations, followed by reattachment at  $x = -0.65$ m on the main plane and  $x = -0.52$ m on the first flap. On the second flap, the flow fully separates without reattachment. These flow features on the wing elements are consistent with the IFW simulations reported by \citet{liosi2024DLES}, with the primary difference being a lower magnitude of pressure peaks in the current case.

\begin{figure}[H]
\centering
\begin{subfigure}{0.6\textwidth}
  \centering
  \includegraphics[width=\linewidth]{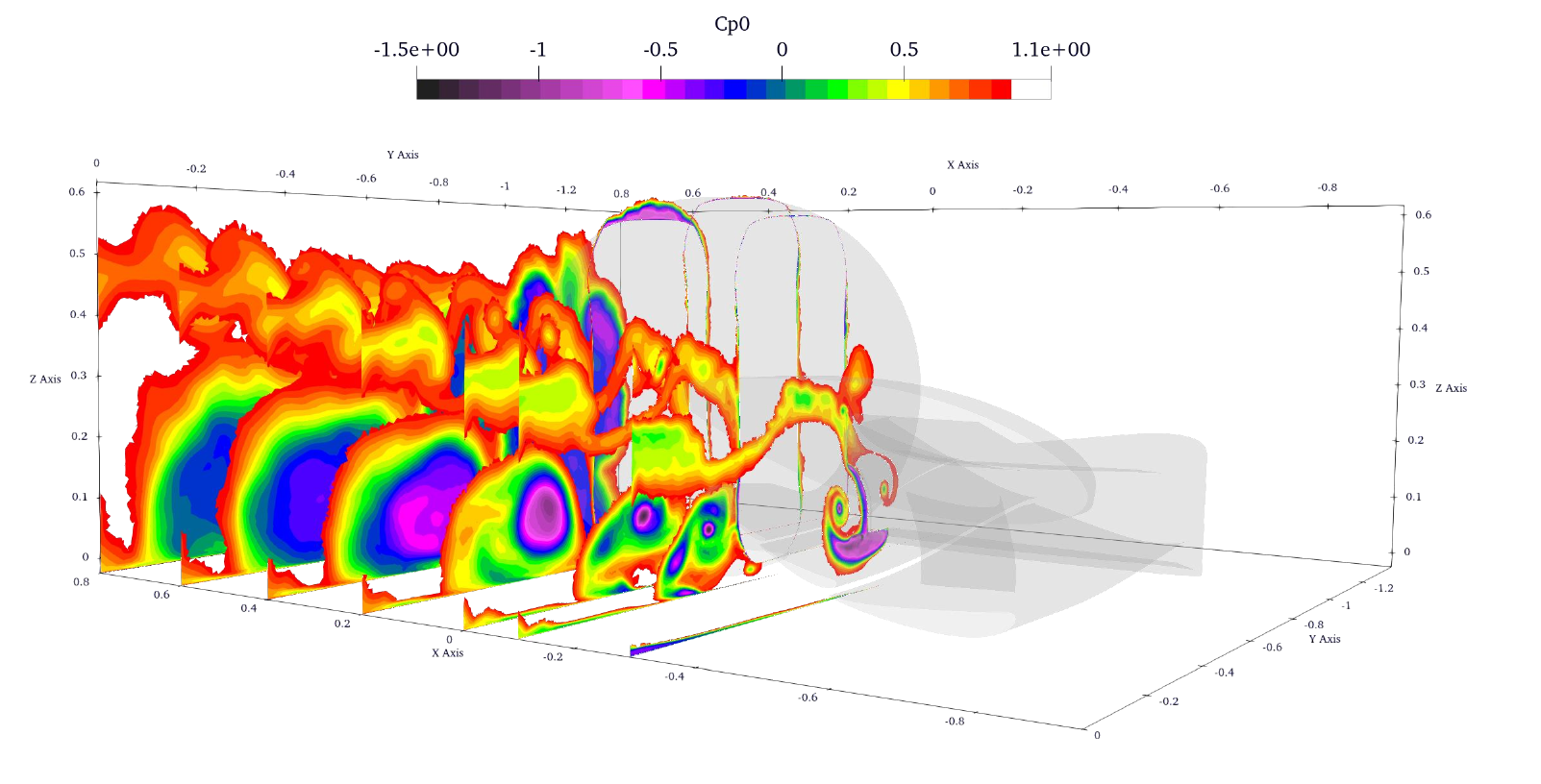}
  \caption{Isometric view}
  \label{subfig:IFWW_xslices_Cp0}
\end{subfigure}%
\begin{subfigure}{0.4\textwidth}
  \centering
  \includegraphics[width=\linewidth]{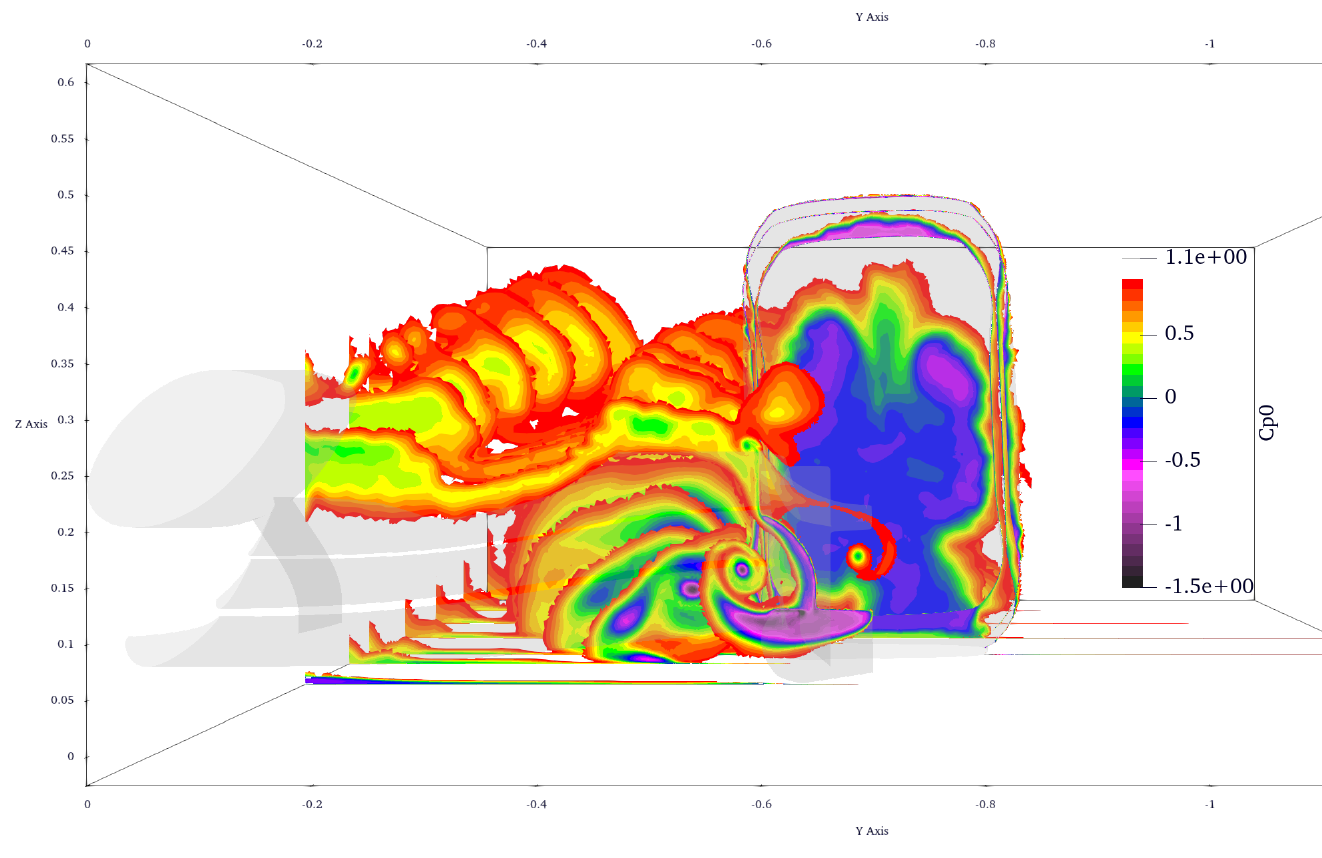}
  \caption{Front view}
  \label{subfig:IFWW_xslices_Cp0_xnorm}
\end{subfigure}
\caption{Time averaged $C_{p0}$ isocontours at different x-slices for the IFW-W simulation.}
\label{fig:IFWW_xslices_Cp0}
\end{figure}

\begin{figure}[H]
    \centering
    \includegraphics[width=\textwidth]{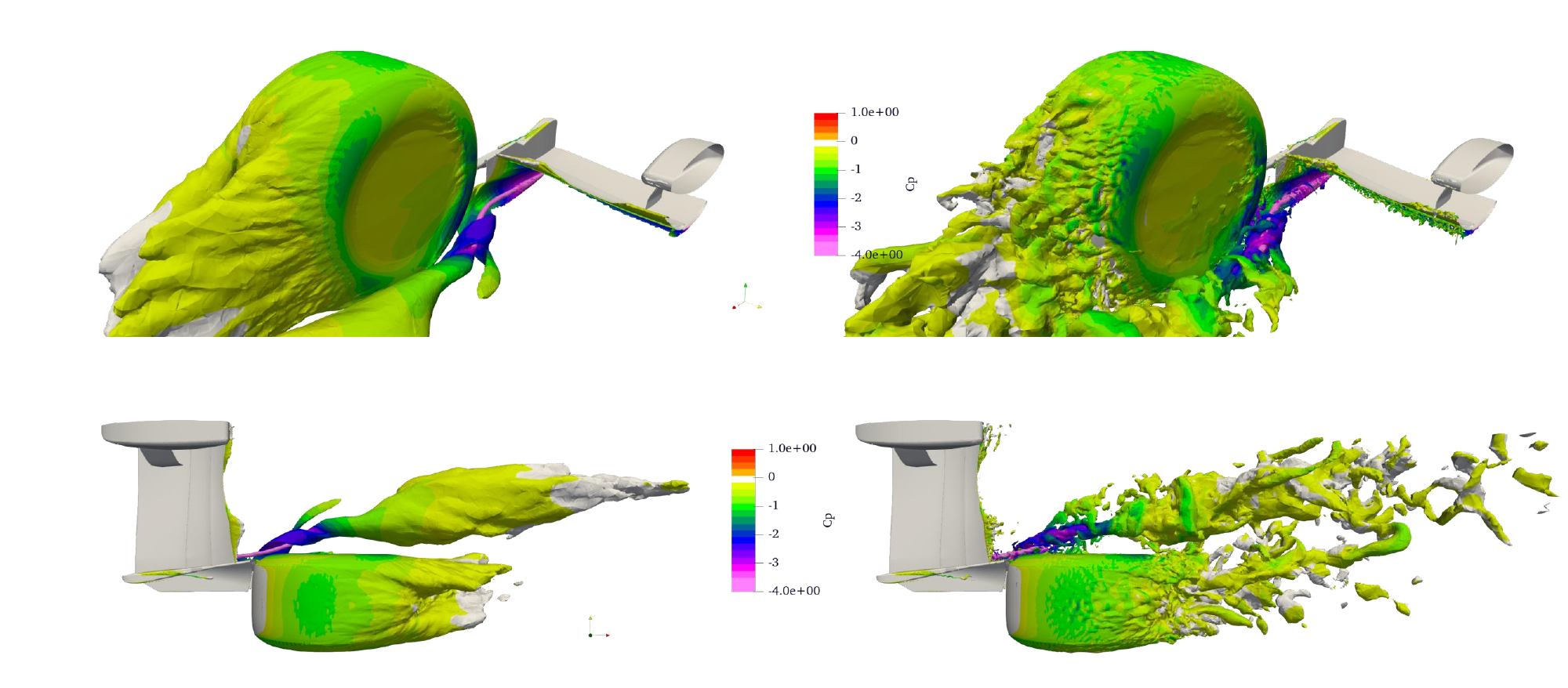} % Replace 'example-image.jpg' with the filename of your picture
    \caption{$C_{p0} = 0$ isocontours colored by $C_p$ for the IFW-W simulation. Left: Time-averaged field for the last 4 CTUs, Right: Instantaneous field at $t^* = 16.5$}
    \label{fig:IFWW_Cp0_iso}
\end{figure}

\autoref{fig:IFWW_xslices_Cp0} illustrates the total pressure coefficient ($C_{p0}$) at various x-locations, providing a clear visualization of the vortex core locations, strength, and trajectory. The isocontours of $C_{p0} = 0$, coloured by the pressure coefficient ($C_p$), shown in \autoref{fig:IFWW_Cp0_iso}, further help visualize the vortex path and identify vortex breakdown. From \autoref{fig:IFWW_xslices_Cp0}, the trajectory of the main vortex is seen starting from the closest x-slice, and as it moves downstream, the vortex rotates, with its trajectory swivelling due to the influence of the wheel. There is also a noticeable interaction with the footplate vortex in the slices closer to the wing. A vortex pair, located at the top of the wheel, is seen in the slice downstream of the wheel in \autoref{subfig:IFWW_xslices_Cp0_xnorm} and is more clearly visible in the average field of \autoref{fig:IFWW_Cp0_iso}. The total pressure contour is qualitatively similar to the results presented by \citet{bearman1988effect} for the wheel wake. \autoref{fig:IFWW_Cp0_iso} also tracks the main vortex's path, showing a significant deflection due to the wheel's influence, with the average field indicating the main vortex bursts on interacting with the wheel. Instantaneous isocontours in \autoref{fig:IFWW_Cp0_iso} highlight the interaction between the wheel and the main vortex and its eventual breakdown. The separation on top of the wheel is seen on top of the wheel, but clear vortices are not identified. 
\subsection{Qualitative comparison of the isolated wheel simulation with the wheel in IFW-W simulations}
The physics around the wheel in the IFW-W simulation from the previous section is inadequately understood. The wake produced by the wheel is a complex phenomenon, and limited published literature addresses this problem.  In this section, an attempt is made to compare qualitatively the physics of a simple isolated wheel to justify the observations around the wheel in the IFW-W geometry.

\begin{figure}[H]
\centering
\begin{subfigure}[t]{0.5\linewidth}
    \centering
    \raisebox{-3.0cm}{% Adjust the vertical position
    \includegraphics[width=\linewidth]{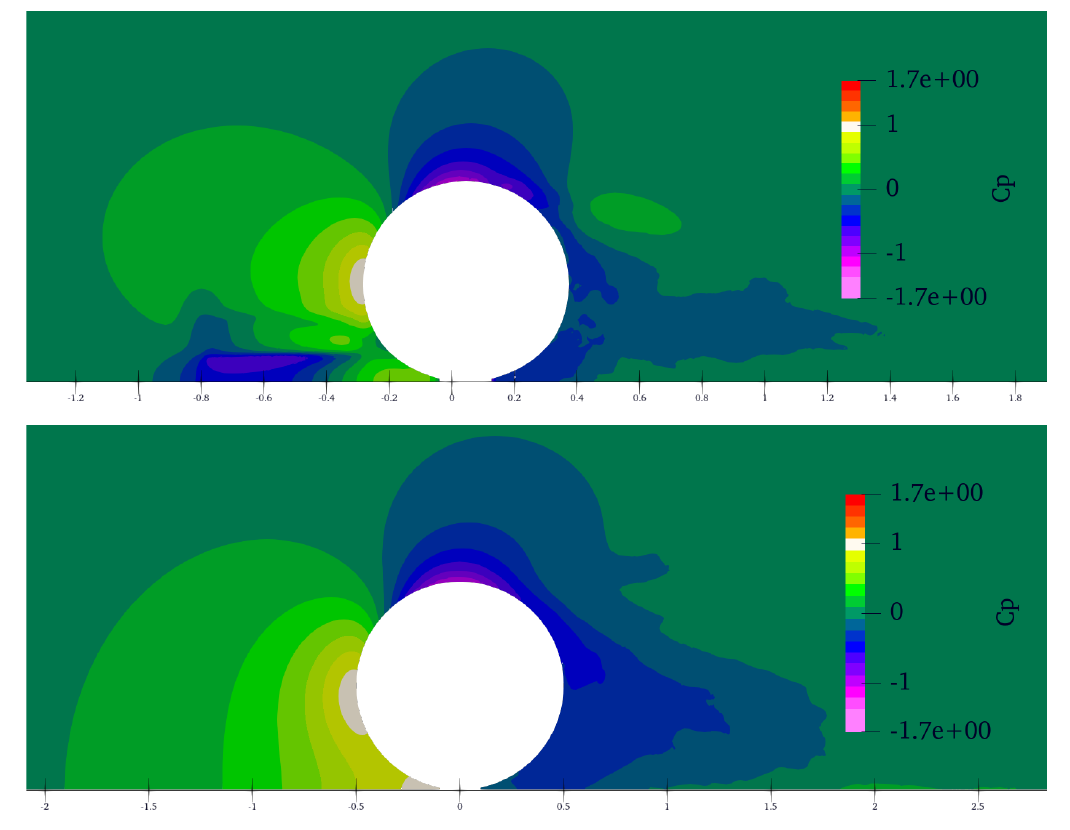}
    }
    \caption{Average Pressure Coefficient ($C_p$) at a slice through the centre of the wheel. Top: IFW-W at y = - 0.7118 m, Bottom: Isolated wheel at y = - 5 m}
    \label{subfig:wheel_Cp}
\end{subfigure}%
\begin{subfigure}[t]{0.5\linewidth}
    \centering
    \begin{minipage}[t]{\linewidth}
        \centering
        \includegraphics[width=\linewidth]{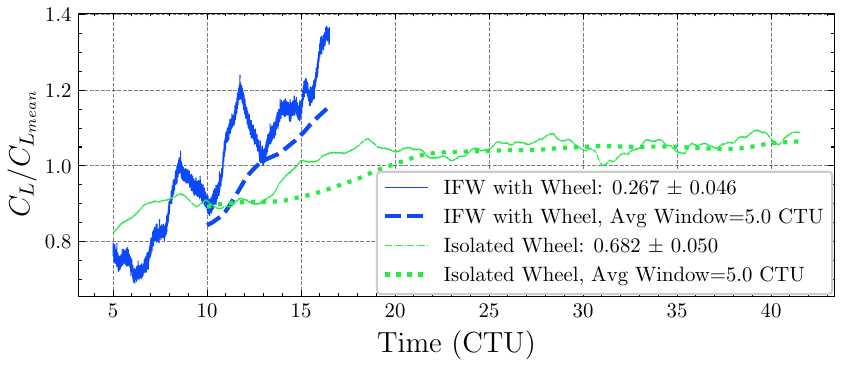}
        \caption{Normalised Lift Coefficient $C_L / C_{L_{mean}}$ vs time}
        \label{subfig:wheel_CL}
    \end{minipage}
    \vspace{0.5cm} % Adjust spacing
    \begin{minipage}[t]{\linewidth}
        \centering
        \includegraphics[width=\linewidth]{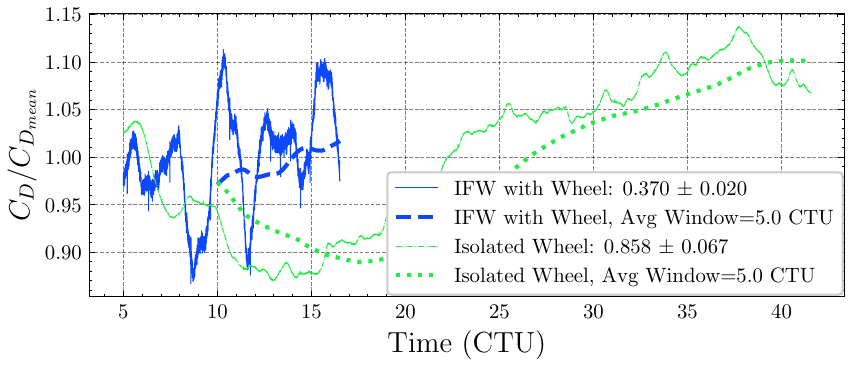}
        \caption{Normalised Drag Coefficient $C_D / C_{D_{mean}}$ vs time}
        \label{subfig:wheel_CD}
    \end{minipage}
\end{subfigure}
\caption{Comparison for the wheel in IFW-W and the isolated wheel simulations. The IFW-W is run with a  $P=4,3$, and the wheel is run using a $P=5,4$}
\label{fig:Wheel_comparison}
\end{figure}

The simulations are run for a simple isolated wheel at the same $Re$, with a polynomial order of $P=5,4$. The simulation is run for 41 CTUs, as seen in the force traces in \autoref{fig:Wheel_comparison}. The simulation results have been time-averaged for the last 24 CTUs. Despite the simulation running for much longer than the IFW-W case in the previous section, it is evident that it has not yet fully achieved statistical convergence. This is corroborated by the high standard deviation ($\geq $5 \%) of the $C_L$ and $C_D$ values and the behaviour of the $C_D$ evolution in \autoref{subfig:wheel_CD}. \autoref{fig:Wheel_comparison} also displays the force traces for the wheel in the IFW-W configuration, highlighting significant standard deviation for both lift and drag.  

The authors assert that additional simulation time is required for the wheel case to reach statistical convergence, as the characteristic length scales involved in both wheel cases are significantly larger than the chosen $L_c$ for the IFW-W geometry. Specifically, the length scale selected for the wheel is the diameter of the wheel, which is $D = 1$ m for the isolated wheel and $D = 0.66$ m for the wheel in the IFW-W; both greater than $L_c = 0.25$ m for the IFW-W. However, running these simulations for extended durations to account for larger characteristic length scales of the wheel demands significantly more computational resources and time. In an industrial context, understanding the required level of fidelity from a simulation is crucial, as it informs decisions on balancing computational cost with the accuracy needed for further action.

\begin{figure}[H]
    \centering
    \includegraphics[width=0.7\textwidth]{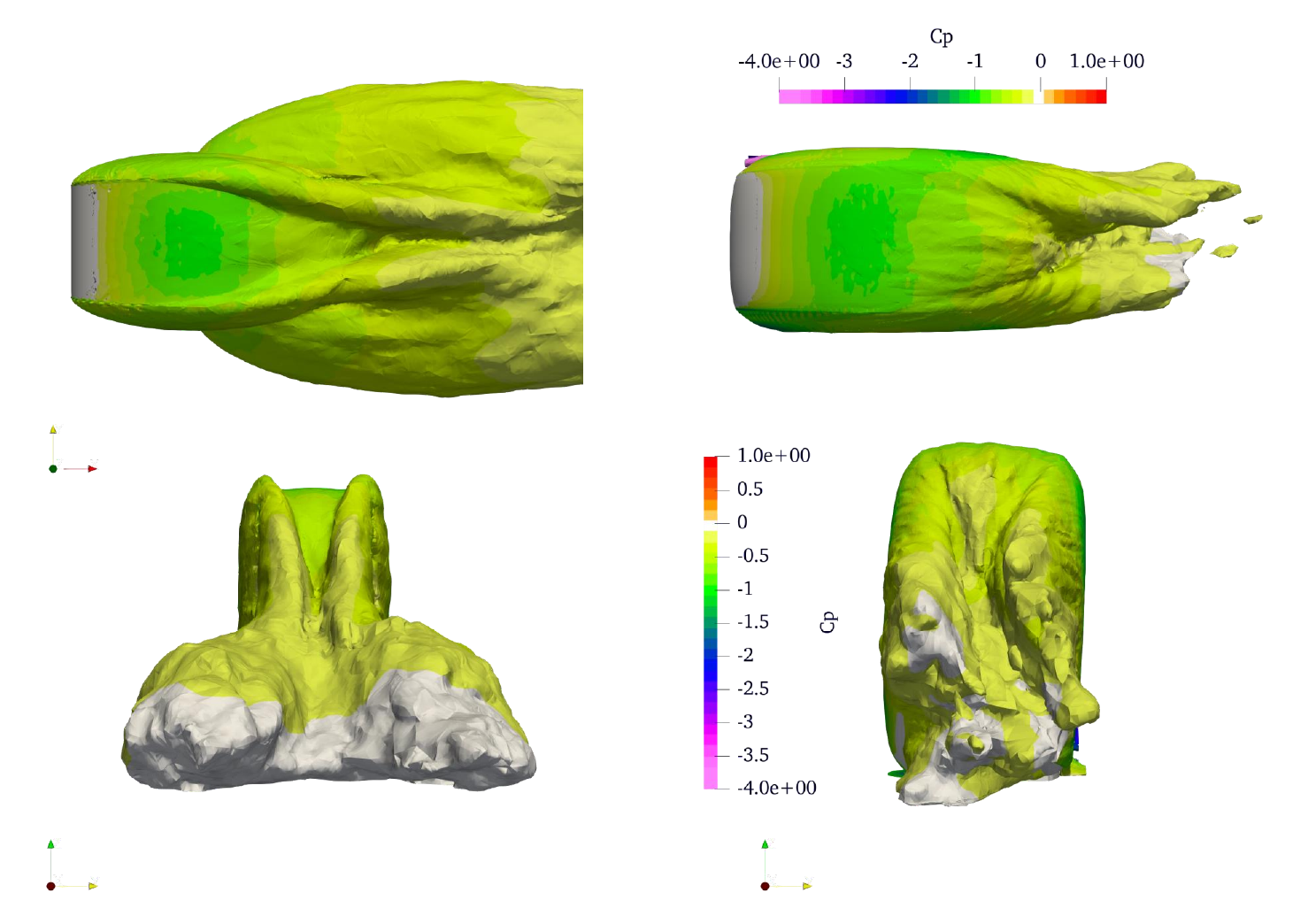} % Replace 'example-image.jpg' with the filename of your picture
    \caption{Time-averaged $C_{p0} = 0$ isocontours colored by $C_p$. The wheel is compared for the IFW-W and the isolated wheel simulations. Left: isolated wheel, Right: IFW-W clipped around the wheel}
    \label{fig:Wheel_Cp0_iso0_comparison}
\end{figure}
Despite the limitations in achieving full statistical convergence, the flow phenomena around both wheel cases exhibit notable qualitative similarities. \autoref{subfig:wheel_Cp} shows the pressure coefficient $C_p$ isocontours for time-averaged solutions on a slice in the middle of the wheel (noting that the wheel in IFW-W has a slight camber, so this is not precisely the centerline). The figure highlights a stagnation region in front of the wheel that extends towards the contact patch. Progressing upward along the wheel circumference, the pressure transitions to the free-stream level ($C_p = 0$) and is followed by regions of negative $C_p$, indicating flow acceleration.

\autoref{fig:Wheel_Cp0_iso0_comparison} illustrates  the $C_{p0} = 0$ isocontours colored by $C_p$ for both wheel cases. Three pairs of counter-rotating vortices were proposed by \citet{mercker1992simulation} and were identified previously in near-turbulent but low $Re=1000$ cases \citep{Pirozzoli2012}. The top pair is visible in the isocontours for both wheels. As proposed by \citet{fackrell1974aerodynamics}, the jetting vortices at the bottom can be seen for the isolated wheel case but are not visible for the IFW-W. The vortex system for the wheel in the IFW-W geometry is distorted due to the flow inboard of the wheel. A more detailed analysis is necessary to understand better the behaviour and trajectory of these vortices in the given configurations. 

%% file: conclusion.tex
\section{Conclusion and Future Work}
This study has demonstrated the industrial application of the spectral/hp element method for simulating incompressible, transitional flows over complex Formula 1 geometries. By conducting simulations of the IFW-W geometry at a moderate Reynolds number of $Re = 2.2 \times 10^5$, the research highlights the feasibility of using high-order methods for industrially relevant aerodynamic analysis. A comprehensive workflow was presented, encompassing key aspects such as meshing strategies, solver configuration, and boundary condition application. Practical guidance on running these simulations, including strategy for convergence assessment, was provided, addressing challenges commonly faced in industrial CFD.

Furthermore, this work introduced a framework for evaluating the fidelity of simulation results in an industrial context. The analysis of the wheel case served as a practical example, illustrating how to interpret and extract meaningful insights from statistically under-converged simulations. The study also emphasized the importance of tailoring the resolution and fidelity requirements to the specific needs of the simulation, leveraging the under-resolved DNS/implicit LES approach to balance computational cost with accuracy.

The present study opens several avenues for further investigation and improvement. A deeper analysis of the IFW-W configuration is needed to better understand the wheel's influence on the overall flow field. Running the simulation for longer durations can provide more statistically converged results for the wheel. Transient flow analysis presents another important direction, particularly for identifying dominant flow frequencies and correlating them to specific flow phenomena. From a computational efficiency perspective, investigating alternate simulation restart strategies can accelerate convergence. Time-stepping methods like substepping \cite{liosi2024DLES} or implicit \cite{henrik2024DLES} approaches, coupled with improved pressure preconditioning, may further enhance the computational efficiency and stability of the solver. Finally, scaling up these simulations to a full-car geometry represents a significant step toward industrial application.

\section*{Acknowledgments}
\begin{minipage}[h]{20mm}
\includegraphics[width=16mm]{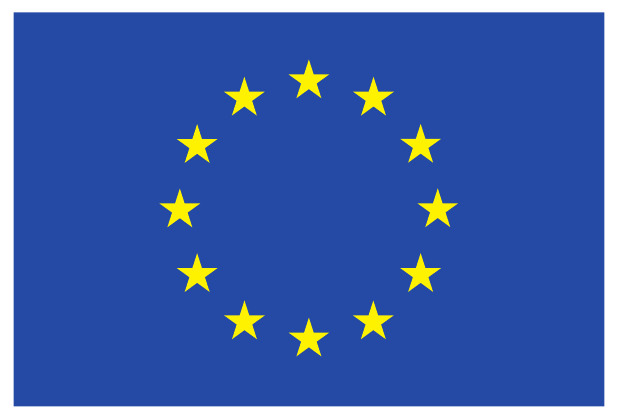}
\end{minipage}\hfill
\begin{minipage}[h]{140mm}
This project received funding from the European Union’s Horizon 2020 research and innovation programme under the Marie Skłodowska-Curie grant agreement No 955923.
\end{minipage} \hfill
This work primarily used the ARCHER2 UK National Supercomputing Service (\url{https://www.archer2.ac.uk}) via the UK Turbulence Consortium (EP/R029326/1). The authors also acknowledge computational resources and support provided by the Imperial College Research Computing Service (\url{http://doi.org/10.14469/hpc/2232}).